\documentclass[namedreferences]{SolarPhysics}
\usepackage{graphicx}        
\usepackage{enumerate}
\usepackage{color}           
\usepackage{url}             
\usepackage {lscape}

\begin{document}
\begin{article}
\begin{opening}

\title{Determination of electron density and filling factor for soft X-ray flare kernels}


\author{J. Jakimiec, U. B\c ak-St\c e\' slicka}

%
\institute{Astronomical Institute, University of Wroc{\l}aw, ul. Kopernika 11, 51-622 Wroc{\l}aw, Poland
                     email: \url{jjakim,bak@astro.uni.wroc.pl}}


\begin{abstract}
In a standard method of determining electron density for soft X-ray (SXR) flare kernels it is necessary to assume what is the extension of a kernel along the line of sight. This is a source of significant uncertainty of the obtained densities.

In our previous paper \cite{bak2005} we have worked out another method of deriving electron density, in which it is not necessary to assume what is the extension of a kernel along the line of sight. The point is that many flares, during their decay phase, evolve along the sequence of steady-state models [quasi-steady-state (QSS) evolution] and then the scaling law, derived for steady-state models, can be used to determine the electron density.

The aim of the present paper is: (1) to improve the two methods of density determination, (2) to compare the densities obtained with the two methods. We have selected a number of flares which showed QSS evolution during the decay phase. For these flares the electron density, $N$, has been derived by means of standard method and with our QSS method. Comparison of the $N$ values obtained with the two different methods allowed us: (1) to test the obtained densities, (2) to evaluate the volume filling factor of the SXR emitting plasma.

Generally, we have found good agreement (no large systematic difference) between the values of electron density obtained with the two methods, but for some cases the values can differ by a factor up to $2$. For most flare kernels estimated filling factor turned out to be about 1, near the flare maximum. 

\end{abstract}

%
\end{opening}

\section{Introduction}
Loop--top flare kernels are the main sources of Soft X-ray (SXR) emission near flares maximum and during their decay. In order to investigate energy balance of the kernels it is necessary to have reliable estimates of plasma density inside the kernels. A "standard` method of determining the electron number density, $N$, consists in a few steps:
\begin{enumerate}[(1)]
\item Usually, the temperature, $T$, is estimated from the ratio of SXR fluxes measured with two different filters:
\begin{equation}
\frac{f_{1}(\bar{T})}{f_{2}(\bar{T})}=\frac{F_{1}}{F_{2}}\\ 
\end{equation}
where $F_{1}$ and $F_{2}$ are measured X--ray fluxes, $f_{i}(T)$ are response functions for measurements with two filters (i=1,2) and $\bar{T}$ is the estimated value of the temperature. In the case of observations with the {\em Reuven Ramaty High Energy Solar Spectroscopic Imager} (RHESSI, \opencite{lin2002}) the temperature is estimated from the slope of SXR spectrum.

Flare kernels are multithermal (non-isothermal), {\em i.e}. 
\begin{equation}
F_{i}=\int \varphi(T) f_{i}(T)dT\\ 
\end{equation}
where $\varphi(T)$ is differential emission measure (DEM). Therefore obtained values of $\bar{T}$ depend on instrumental response functions $f_{i}(T)$ and we obtain different values of $\bar{T}$ from measurements with different instruments (see Table \ref{table}).
\item Next the emission measure is calculated from the formula:
\begin{equation}
\label{EM}
EM=F_{i}/f_{i}(\bar T)\\ 
\end{equation}

In many cases the response function, $f_{i} (T)$, steeply depends on temperature. Then random errors of determination of mean temperature, $\bar T$, may cause significant errors of $EM$. To improve this standard method of $EM$ determination, we recommend that observations having flat response functions (low $|df_{i}$/$dT|$ in the investigated range of temperatures) should be used to determine $EM$ from Equation (\ref{EM}) (see Section 2). Then we have:
\begin{equation}
\label {fi}
F_{i}=\int\varphi(T) f_{i}(T) dT\approx \bar f_{i}\int\varphi(T) dT=\bar f_{i}~EM,\\ 
\end{equation}
{\em i.e.} we will obtain reliable $EM$ from Equation (\ref{EM}), despite that the emitting plasma is multithermal.
\item The electron number density is estimated as:
\begin{equation}
\label {N}
N=\sqrt{EM/V},\\ 
\end{equation}
where $V$ is the volume of emitting plasma. In order to estimate the volume $V$ it is necessary to assume what is the extension of emitting plasma along the line of sight. This introduces significant uncertainty into estimates of electron density from Equation~(\ref{N}). The uncertainty is high when the shape of a flare kernel is elongated (elliptical) or irregular, since then it is most difficult to guess what is the extension of emitting plasma along the line of sight.

Additional drawback of standard method of density determination is due to the fact that it gives values averaged over whole volume of a kernel. In reality, the emitting plasma can have higher density if it occupies only a part of kernel volume, {\em i.e.} if its filling factor is smaller than 1.\\
\end{enumerate}
In our previous paper (\inlinecite {bak2005}, hereafter Paper I) we have proposed a new method of electron density determination in which (1) it is not necessary to assume what is the extension of the emitting plasma along the line of sight and (2) which gives actual values of the electron density of emitting plasma.\\
\inlinecite{rosner78} investigated models of steady-state coronal loops and they have found a ``scaling law'' of these models which can be written in the form:
\begin{equation}
\label {rosner}
N_{SS} = 2.1\times 10^{6}T^{2}/L
\end{equation}
where $T$ is the temperature at the top of a loop, $L$ is the semilength of the loop and $N_{SS}$ is the electron number density within the loop.\\ 
Those authors have found that Equation (\ref{rosner}) is also valid for hot ``post-flare loops'' having temperature up to 10 MK (see their Figure 9b). This last finding has been explained by \inlinecite{jakimiec92} who carried out hydrodynamic modeling of the evolution of hot flaring loops. They have found that if the heating rate, $E_{H}(t)$, of a flaring loop decreases slowly during decay phase, the loop evolves along a sequence of steady-state models [quasi-steady-state (QSS) evolution] and then Equation (\ref{rosner}) is valid during this evolution. Therefore this Equation (\ref{rosner}) can be used to determine densities $N$ during the QSS evolution.
The aim of the present paper is: (1) to improve the two methods of density determination, (2) to compare the densities obtained with the two methods. We have selected a number of flares which showed QSS evolution during the decay phase. For these flares the electron density, $N$, has been derived by means of standard method and with our QSS method. Comparison of the $N$ values obtained with the two different methods allowed us: (1) to test the obtained densities, (2) to evaluate the volume filling factor of the SXR emitting plasma.

\begin{figure}[!h]
\begin{center}
\vspace{0.1cm}
\includegraphics[width=0.48\textwidth,clip=]{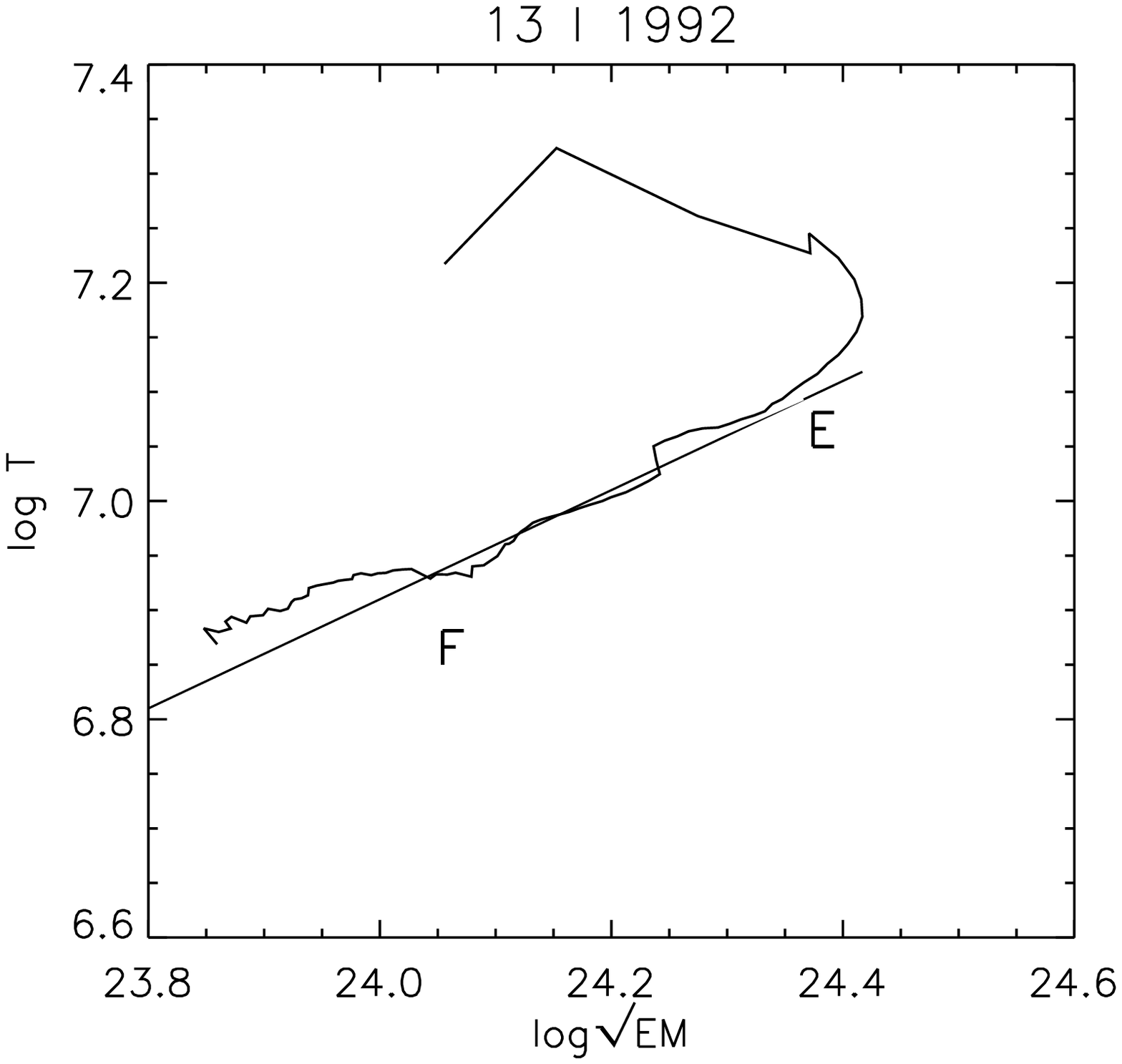}
\includegraphics[width=0.47\textwidth,clip=]{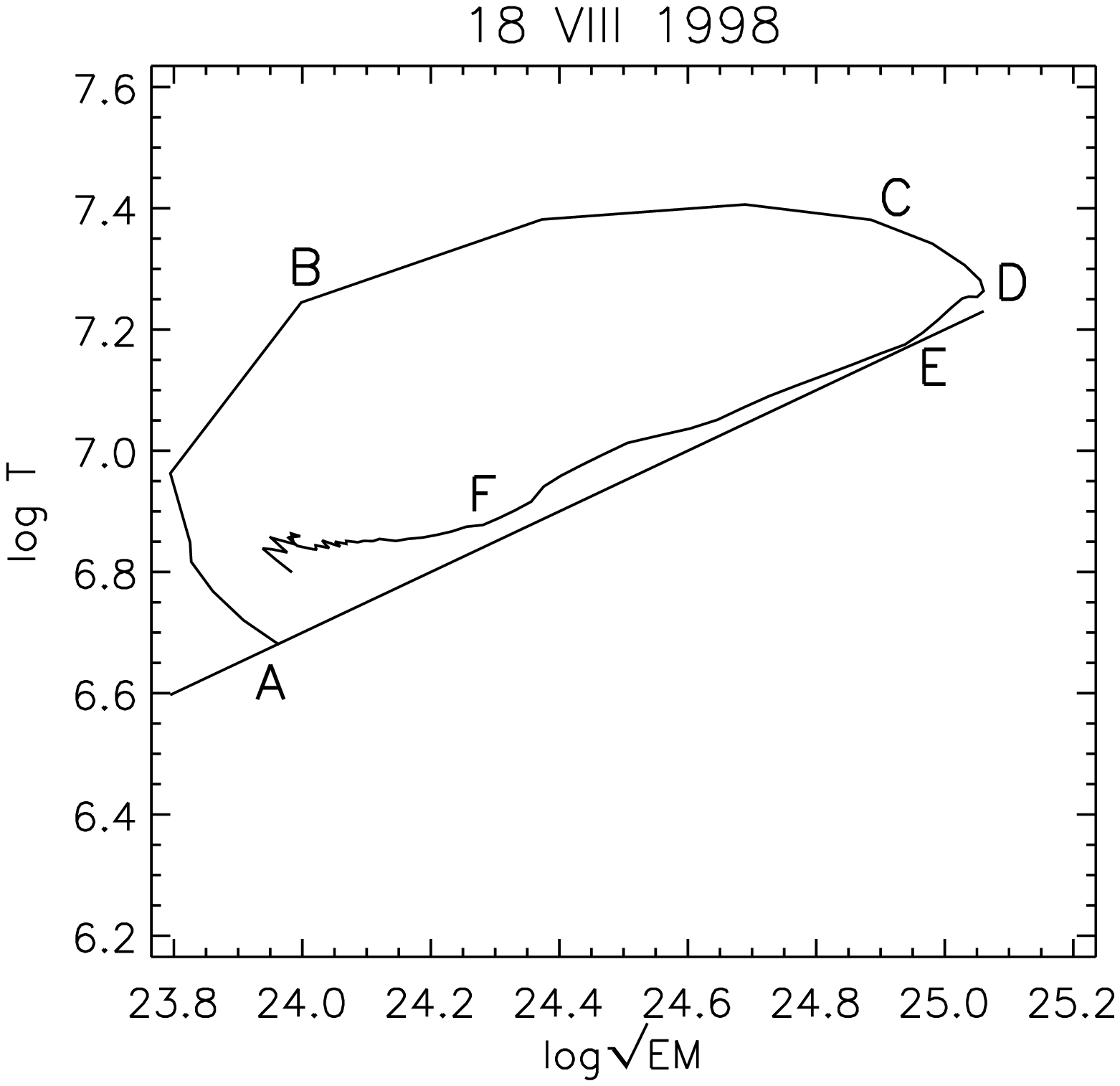}
\vspace{0.3cm}
\includegraphics[width=0.48\textwidth,clip=]{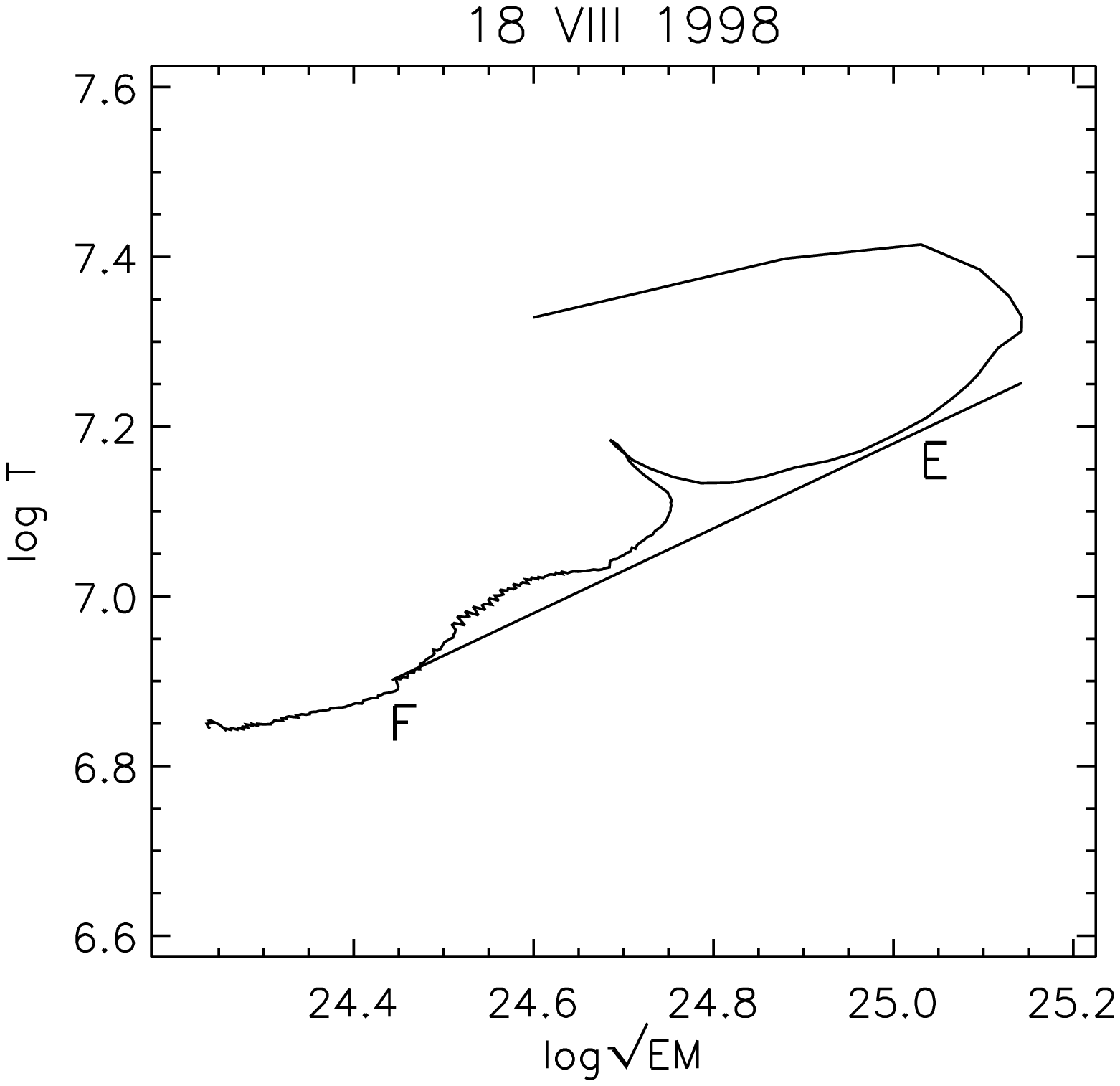}
\includegraphics[width=0.48\textwidth,clip=]{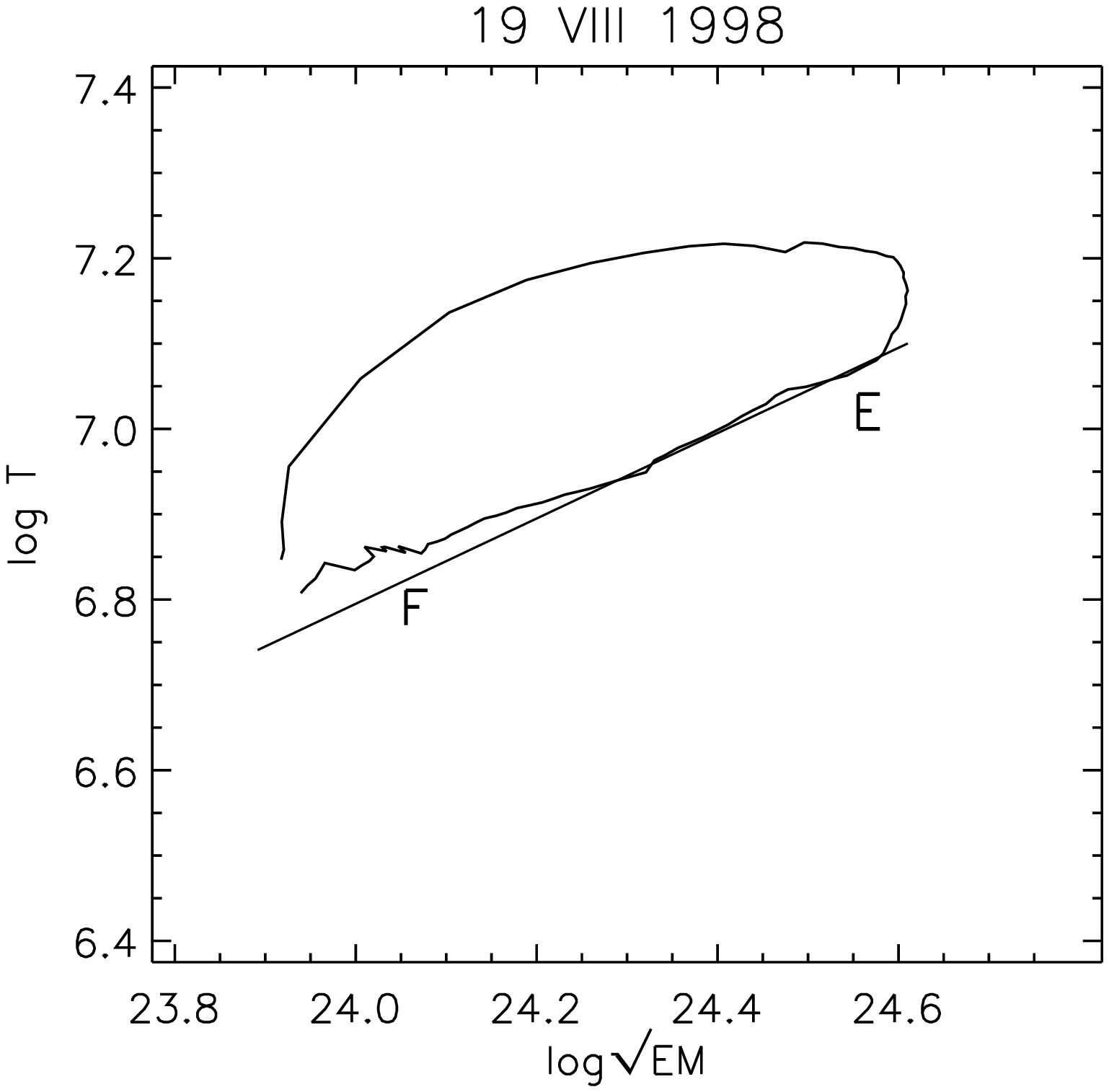}
\vspace{0.3cm}
\includegraphics[width=0.48\textwidth,clip=]{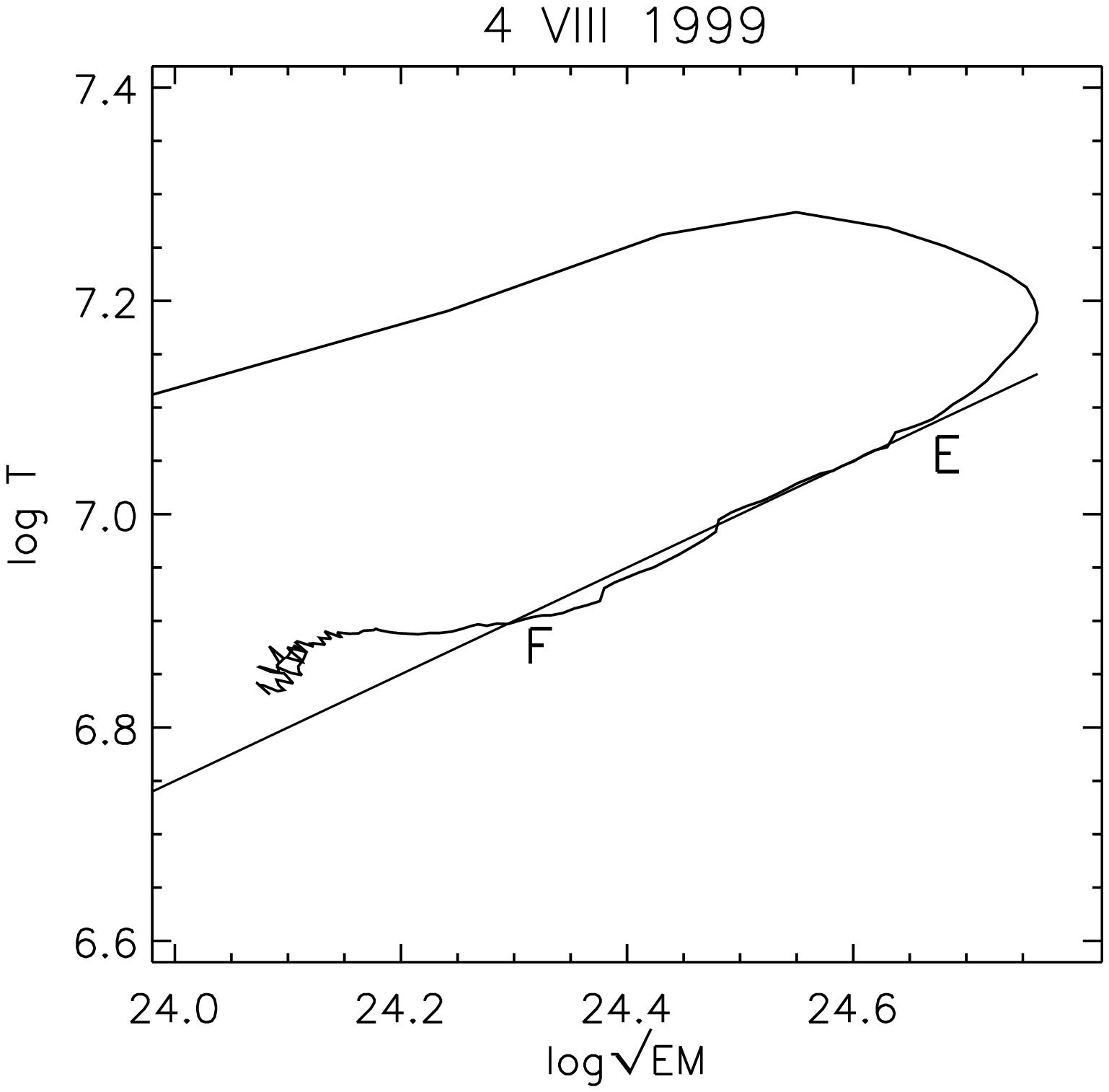}
\includegraphics[width=0.485\textwidth,clip=]{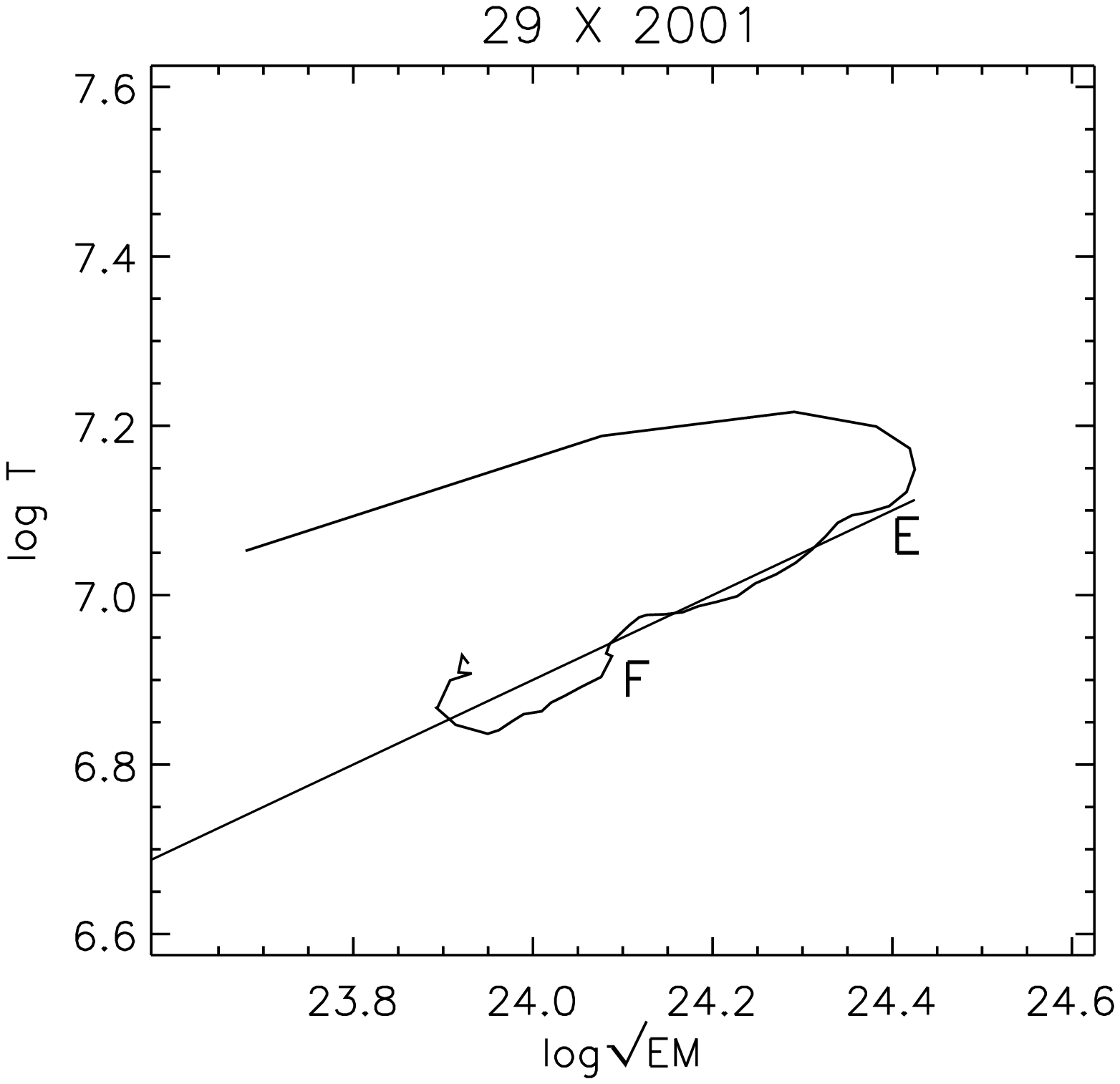}
\caption{Examples of diagnostic diagrams obtained from the GOES data. Flare evolution proceeds from A to F. EF is the QSS branch. Straight lines have inclination $\zeta=0.5$.}
\label{diagn}
\end{center}
\end{figure}

Observations and their analysis are described in Section 2. Section 3 contains summary and conclusions.

\section{Observations and their analysis}

Long time ago it has been shown that it is useful to display flare evolution in temperature-density diagnostic diagrams (\citeauthor{jakimiec87} (\citeyear{jakimiec87, jakimiec92})). In such diagrams the QSS branch of flare evolution is a straight line of inclination $\zeta=dlogT/dlogN\approx0.5$. Soft X-ray images show that size of loop-top sources does not change significantly during decay phase of many flares ({\em i.e} their volume $V\approx const$). Therefore in Paper~I we have found that simplified diagnostic diagrams (where $logN$ is replaced by $log\sqrt EM$) derived from GOES ({\em Geostationary Operational Environmental Satellites}) observations, are very useful for detection of the QSS evolution (see Figure \ref{diagn}). Correct inclination, $\zeta\approx0.5$, of the QSS branch in these diagrams indicates that temperature, $T_{GOES}$, derived from {\em GOES} observations, adequately represents mean temperature of emitting plasma during the QSS flare evolution and therefore we should use this temperature in Equation~(\ref{rosner}).

In Paper I we have also found that temperatures $T_{SXT}$ are usually lower than $T_{GOES}$ (SXT is the {\em Yohkoh}/Soft X-ray Telescope - see \opencite{tsuneta91}). This is caused by the fact that SXT has narrower spectral range of sensitivity ($1$--$5$ keV for Be $+$ Al12 filters; $1.6$--$25$ keV for GOES). Low values of $T_{SXT}$ result in low inclination ($\zeta<0.5$) of the QSS branch derived with this temperature (see \inlinecite{kolomanski02} and our Paper I).

\begin{figure}[!h]
\begin{center}
\includegraphics[width=12cm]{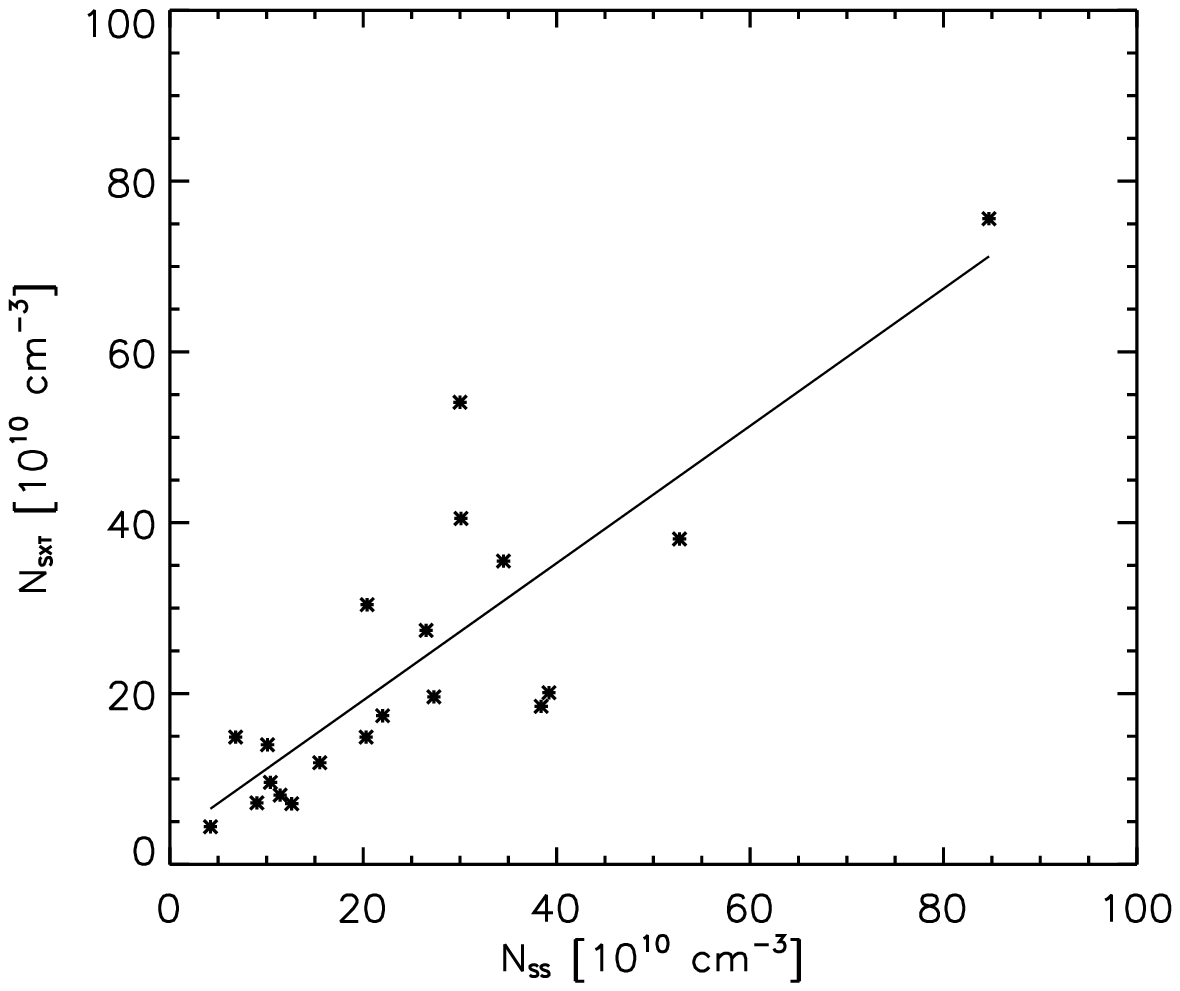}
\caption{Comparison of $N_{SXT}$ and $N_{SS}$ for $20$ selected flares. Estimates of random errors are given in Table 1.}
\label{NsxtNss}
\end{center}
\end{figure}

\subsection{Comparison of the electron densities obtained with two independent methods}
\begin{figure}[!t]
\begin{center}
\includegraphics[width=10cm]{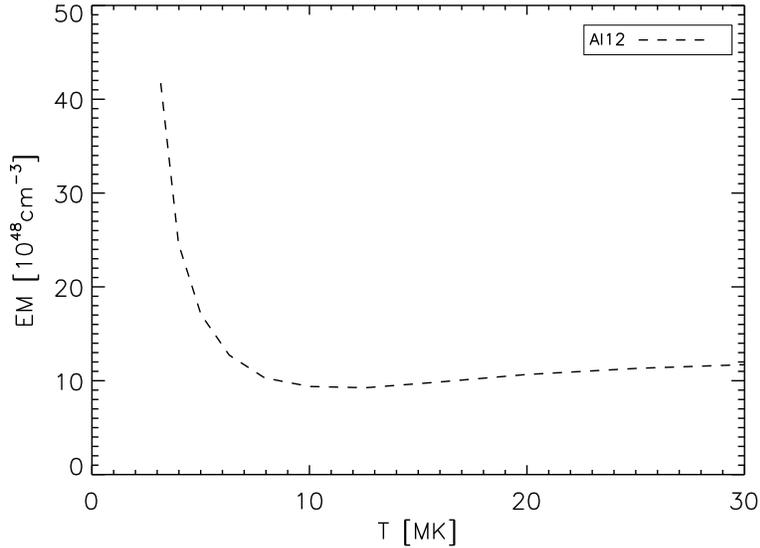}
\caption{$EM$ for various values of temperature. The calculations have been carried out for a constant value of the SXR flux, $F_{i}$, measured with the Al12 filter ($F_{i}=1.5\times10^{6}$ DN).}
 \label{EMvsT}
\end{center}
\end{figure}

Using {\em Yohkoh}/SXT and GOES observations we have selected $20$ flares which are limb or near-the-limb flares and their GOES diagnostic diagrams show clear QSS branch. For the beginning of the QSS branch (near point E in Figure \ref{diagn}) we have determined temperature, $T_{SXT}$ and $T_{GOES}$, and densities, $N_{SXT}$ and $N_{SS}$, using the SXT images and GOES data. The flaring loop semilength, $L$, has been calculated as $L=\frac{\pi}{2}h$, where $h$ is the loop height measured in the SXR image. Obtained values of the temperatures and densities are given in Table \ref{table}. Comparison of $N_{SXT}$ and $N_{SS}$ is shown in Figure \ref{NsxtNss}. We see good agreement between $N_{SXT}$ and $N_{SS}$ values. The regression line is: $N_{SXT} = 0.803 N_{SS} + 3.13$ [this line is $N_{SXT} = 0.784 N_{SS} + 2.21$ if we neglect the point ($30$,$54$) which shows large deviation -- see discussion below]. The correlation coefficient is $r=0.844$.
\begin{landscape}
\begin{table}[!t]
	\caption{List of selected flares}
	\vspace{3mm}
	\label{table}
	\begin{small}
		\begin{tabular}{|c|c|c|c|c|c|c|c|c|c|c|c|}
\hline

&&GOES&GOES&\multicolumn{8}{c|}{Beginning of the QSS evolution}\\
&Date&class&max.&Start&$T_{SXT}$&$T_{GOES}$&$N_{SXT}$&$\Delta N_{SXT}$&$N_{SS}$&$\Delta N_{SS}$&$k$\\
&&&(UT)&(UT)&(MK)&(MK)&&&&&\\
\hline
1.& 1991 Nov. 04 &M3.4&23:23&23:47&8.0&11.5&7.1&1.4&12.6&2.5&0.6\\
2.& 1991 Nov. 19 &C8.5&09:32&09:36&8.0&11.3&9.6&1.9&10.4&2.1&0.9\\
3.& 1991 Dec. 03 &X2.2&16:39&16:40&10.4&19.1&75.6&15.1&84.7&16.9&0.9\\
4.& 1992 Jan. 13 &M2.0&17:34&17:41&8.2&12.8&11.9&2.4&15.5&3.1&0.8\\
5.& 1992 Sep. 09&M3.1&02:13&02:15&9.2&14.0&19.6&3.9&27.3&5.5&0.7\\
6.& 1992 Oct. 04 &M2.4&22:27&22:31&8.1&12.5&14.9&3.0&20.3&4.1&0.7\\
7.& 1992 Nov. 22 &M1.6&23:10&23:11&9.0&13.3&20.1&4.0&39.2&7.8&0.5\\
8.& 1993 Jan. 02 &C8.4&23:52&23:59&7.7&9.1&7.2&1.4&9.0&1.8&0.8\\
9.& 1993 Mar. 23 &M2.3&01:53&02:11&8.3&10.7&4.4&0.9&4.2&0.8&1.0\\
10.& 1994 Jan. 16 &M6.1&23:25&23:35&9.0&12.3&30.4&6.1&20.4&4.1&1.5\\
11.& 1998 Aug. 18 &X2.8&08:24&08:31&10.9&15.6&40.5&8.1&30.1&6.0&1.3\\
12.& 1998 Aug. 18 &X4.5&22:15&22:27&11.1&17.1&54.1&10.8&30.0&6.0&1.8\\
13.& 1998 Aug. 19 &M3.0&14:26&14:50&8.0&10.3&14.9&3.0&6.8&1.4&2.2\\
14.& 1998 Nov. 22 &X3.7&06:42&06:45&11.2&16.5&38.1&7.6&52.7&10.5&0.7\\
15.& 1998 Nov. 22 &X2.5&16:23&16:28&12.3&16.3&35.5&7.1&34.5&6.9&1.0\\
16.& 1998 Nov. 23 &X2.2&06:44&06:57&11.3&15.3&27.4&5.5&26.5&7.3&1.0\\
17.& 1999 Aug. 04 &M6.0&05:57&06:09&8.5&12.9&17.4&3.5&22.0&4.4&0.8\\
18.& 2000 Jan. 12 &M1.1&20:49&20:57&8.1&10.6&8.1&1.6&11.4&2.3&0.7\\
19.& 2000 Jun. 23 &M3.0&14:31&14:42&9.0&12.3&14.0&2.8&10.1&2.0&1.4\\
20.& 2001 Oct. 29 &M1.3&01:59&02:01&8.3&13.2&18.5&3.7&38.4&7.7&0.5\\
\hline
\multicolumn{12}{@{} l @{}}{Values of $N_{SXT}$ and $N_{SS}$ are in $10^{10} cm^{-3}$, $k=N_{SXT}/N_{SS}$, $\Delta N$ are $20\%$ random errors (see text)} \\	
	
	\end{tabular}
	\end{small}
\end{table}
\end{landscape}
The good agreement between $N_{SXT}$ and $N_{SS}$ values shows that both methods of density determination give reliable values of the density (no large systematic errors in their calibration).\\ 
It is surprising that we obtain agreement between the densities despite of significant systematic differences between the temperatures $T_{SXT}$ and $T_{GOES}$. Explanation of this paradox is the following: The response function, $f_{i}(T)$ for the {\em Yohkoh}/SXT with Al12 filter changes only a little in wide range of temperature, $T \approx 8-40$ MK [see curve $f$ in Figure 9 in \inlinecite{tsuneta91}]. Therefore, the emission measure, calculated according to Equation (\ref{EM}), only weakly depends on temperature $\bar T$. To illustrate this effect, we have assumed a constant value of the X-ray flux, $F_{i}$, and calculated $EM$ for various values of temperature (Figure \ref{EMvsT}). For Al12 filter the dependence of $EM$ on $T$ is weak in the range of flare temperatures. Dependence of the electron density, $N_{SXT}$, on temperature is even weaker, since $N_{SXT}\sim \sqrt{EM}$. This explains why we obtain correct values of $N_{SXT}$, despite of the fact that $T_{SXT}$ and $|dT_{SXT}/dt|$ are systematically too low.

Figure \ref{EMvsT} displays function $1.5\times 10^{6} [DN]/f_{i}(T)$ for SXT observations with Al12 filter. We have calculated values of this function for $T_{SXT}$ values which are presented in Table 1. From these values of the function we obtained its mean value and random ({\em r.m.s.}) error. This gives the following form of Equation (\ref{EM}):\\ 
\begin{equation}
\label{EM_F}
EM [10^{48} cm^{-3}] = (6.48 \pm 0.27 )\times 10^{-6}F_{i}[DN]
\end{equation} 
which does not depend on estimated value of mean temperature $\bar T$, provided that $\bar T \geq 8$ MK (DN is Digital Number in {\em Yohkoh}/SXT observations). However, the digital coefficient in Equation (\ref{EM_F}) can slowly change with time due to slow change in calibration of the instrument.

\begin{figure}[!h]
\begin{center}
\includegraphics[width=0.48\textwidth,clip=]{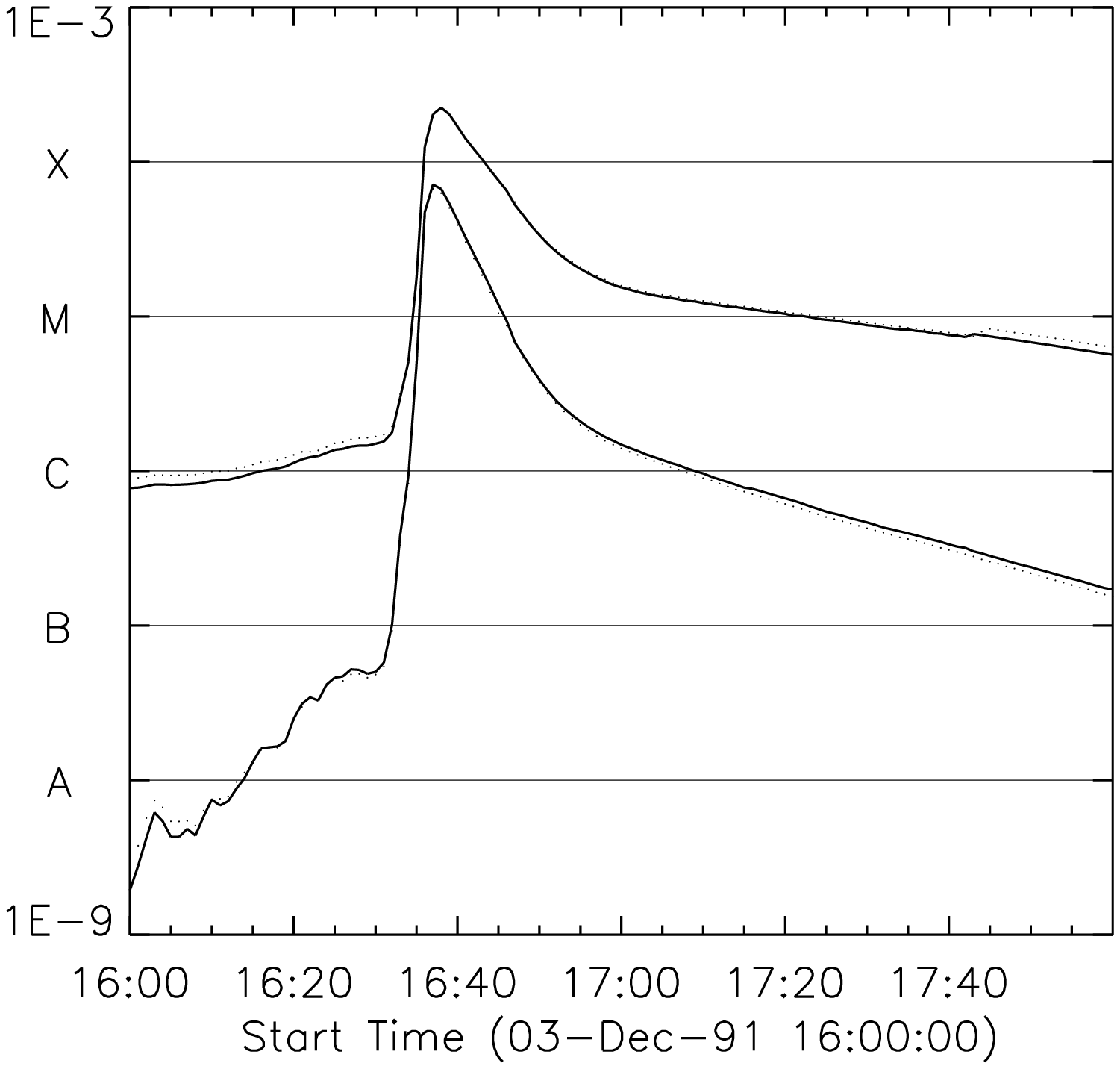}
\includegraphics[width=0.51\textwidth,clip=]{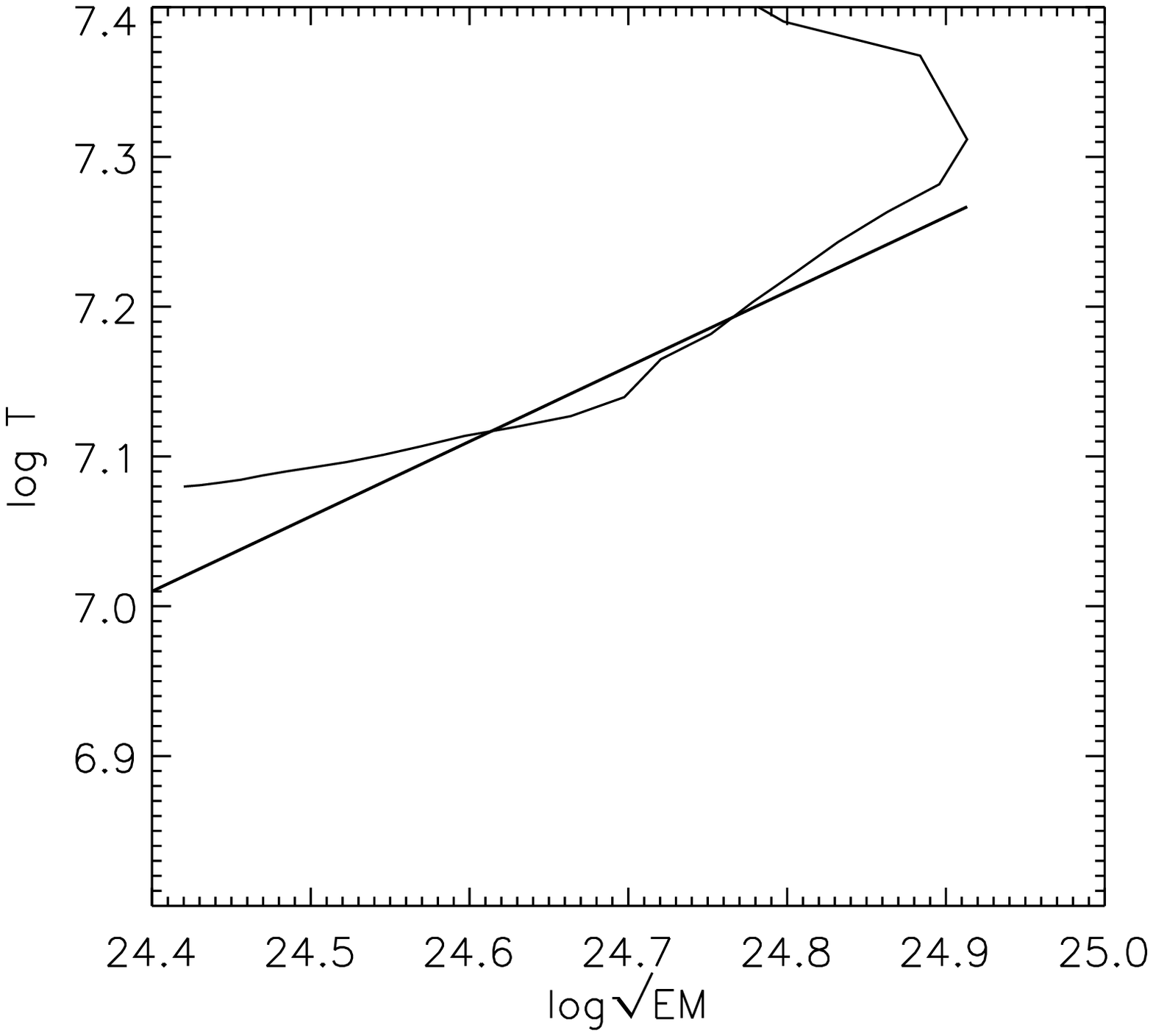}
\includegraphics[width=7cm]{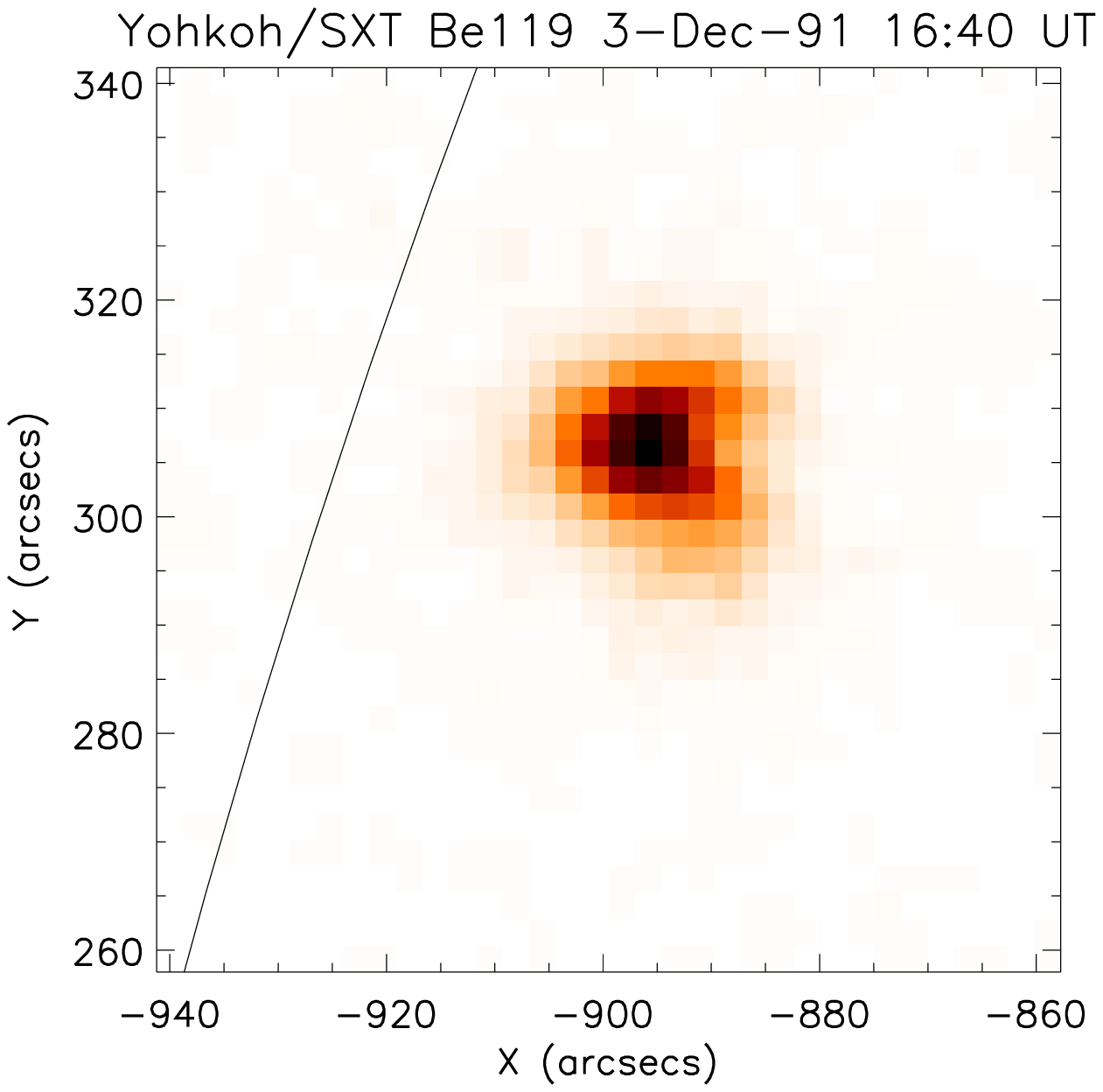}
\caption{GOES light curve, diagnostic diagram and sample SXT(Be119) image (inverse intensity scale in colour) of December 3, 1991 flare (solid line shows solar limb).}
 \label{goes3dec91}
\end{center}
\end{figure}

In Table \ref{table} we see that most flares in our sample were strong ones (GOES class M and X). Electron number density, measured near SXR flare maximum, for most of the flares falls within the limits of ($4-50$) $\times 10^{10}$ cm $^{-3}$. But for one flare [No. 3 in Table \ref{table}, point (85,76) in Figure \ref{NsxtNss}] the density was extremely high, $N \approx 8$ $\times 10^{11}$ cm $^{-3}$. This high density has been obtained with both methods of density determination. SXR image of this flare is shown in Figure~\ref{goes3dec91}. We see that it was a low flaring loop with very bright, compact loop-top kernel (at the beginning of the QSS evolution the diameter of the kernel was about $6000$ km). This flare was also very strong in HXRs during its impulsive phase (about $1300$ counts/s/SC in the {\em Yohkoh}/HXT $23$-$33$ keV light curves, SC is a subcollimator). 

\begin{figure}[!t]
\begin{center}
\includegraphics[width=10cm]{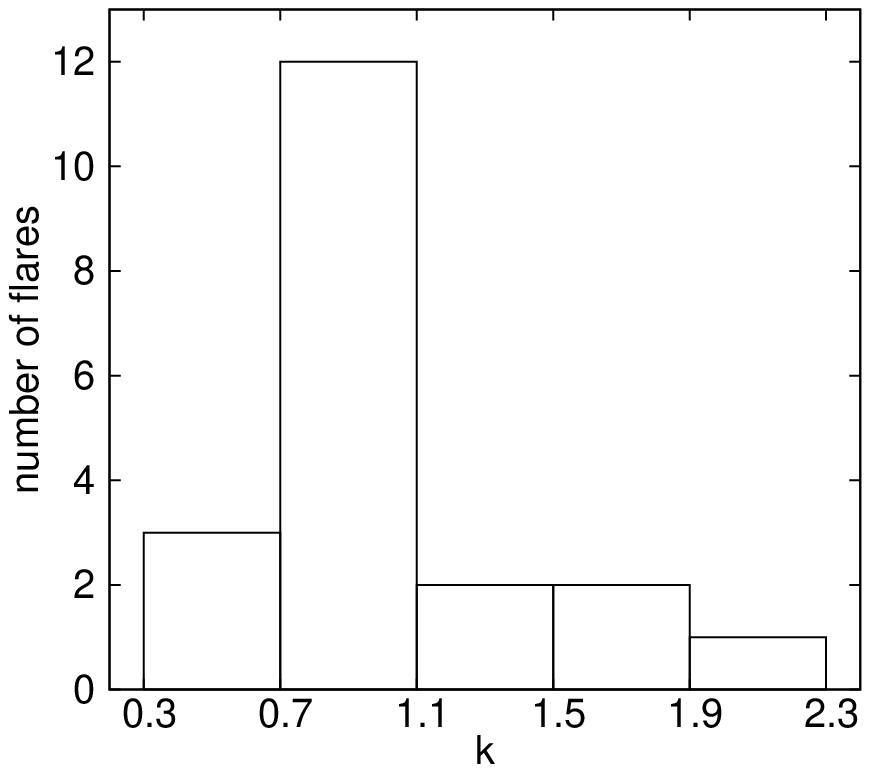}
\caption{Histogram of the ratio $k=N_{SXT}/N_{SS}$ for 20 analysed flares}
 \label{hist}
\end{center}
\end{figure}
Main source of random errors (deviations from the regression line in Figure \ref{NsxtNss}) in the standard method of density determination from SXR images, is due to the fact that we should assume what is extension of a SXR kernel along the line of sight. Main source of random errors in the QSS method of density determination is due to uncertainty in estimates of the flaring-loop semilength, $L$. Estimates of $L$ are usually based on assumption that the shape of the loop is semicircular and (in the case of limb flares) that the loop lies in the plane of image. Deviations from these assumptions are the main source of random errors in the QSS method of electron density determination.
 
\begin{figure}
\begin{center}
\includegraphics[width=6cm]{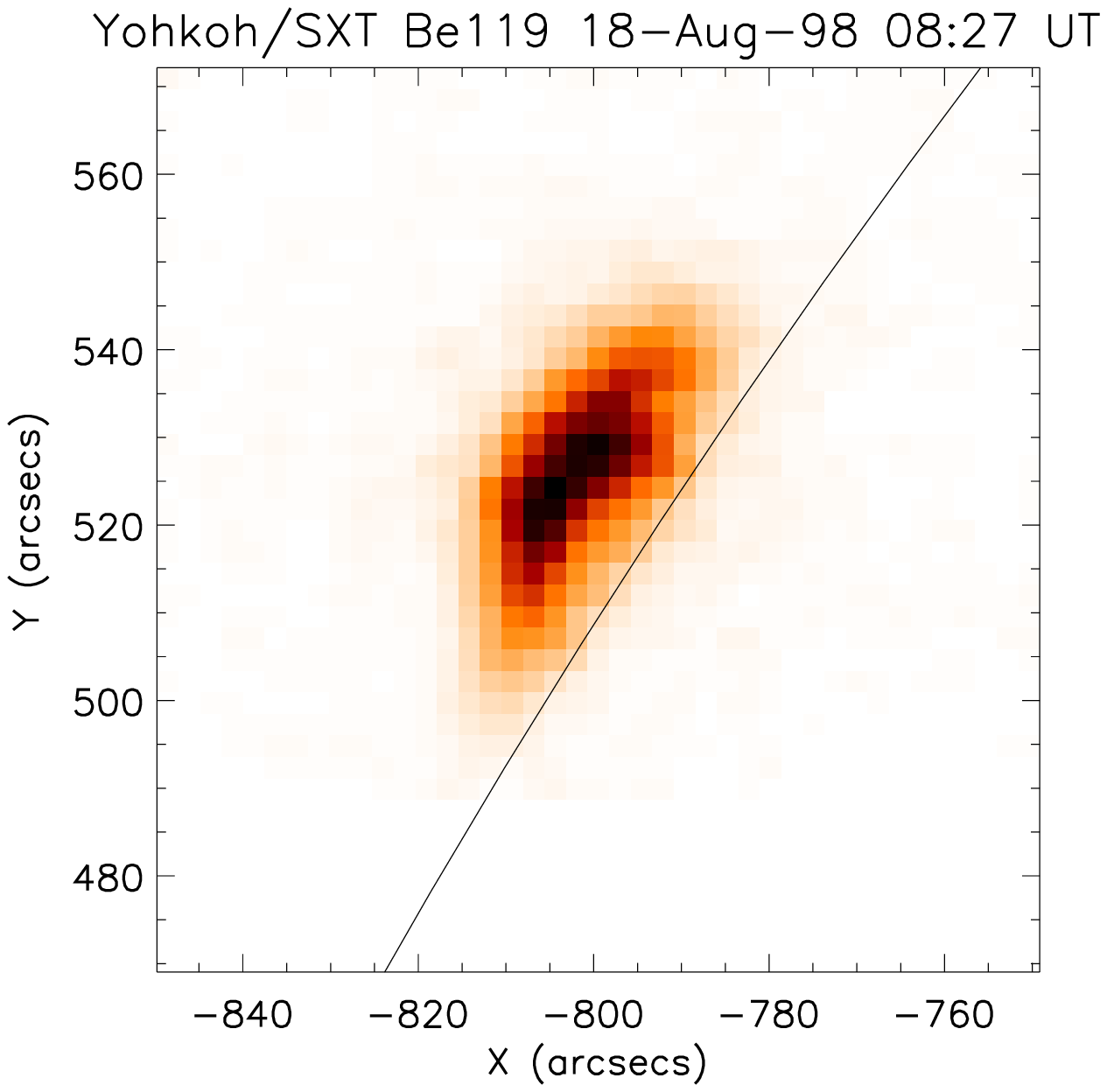}
\includegraphics[width=6cm]{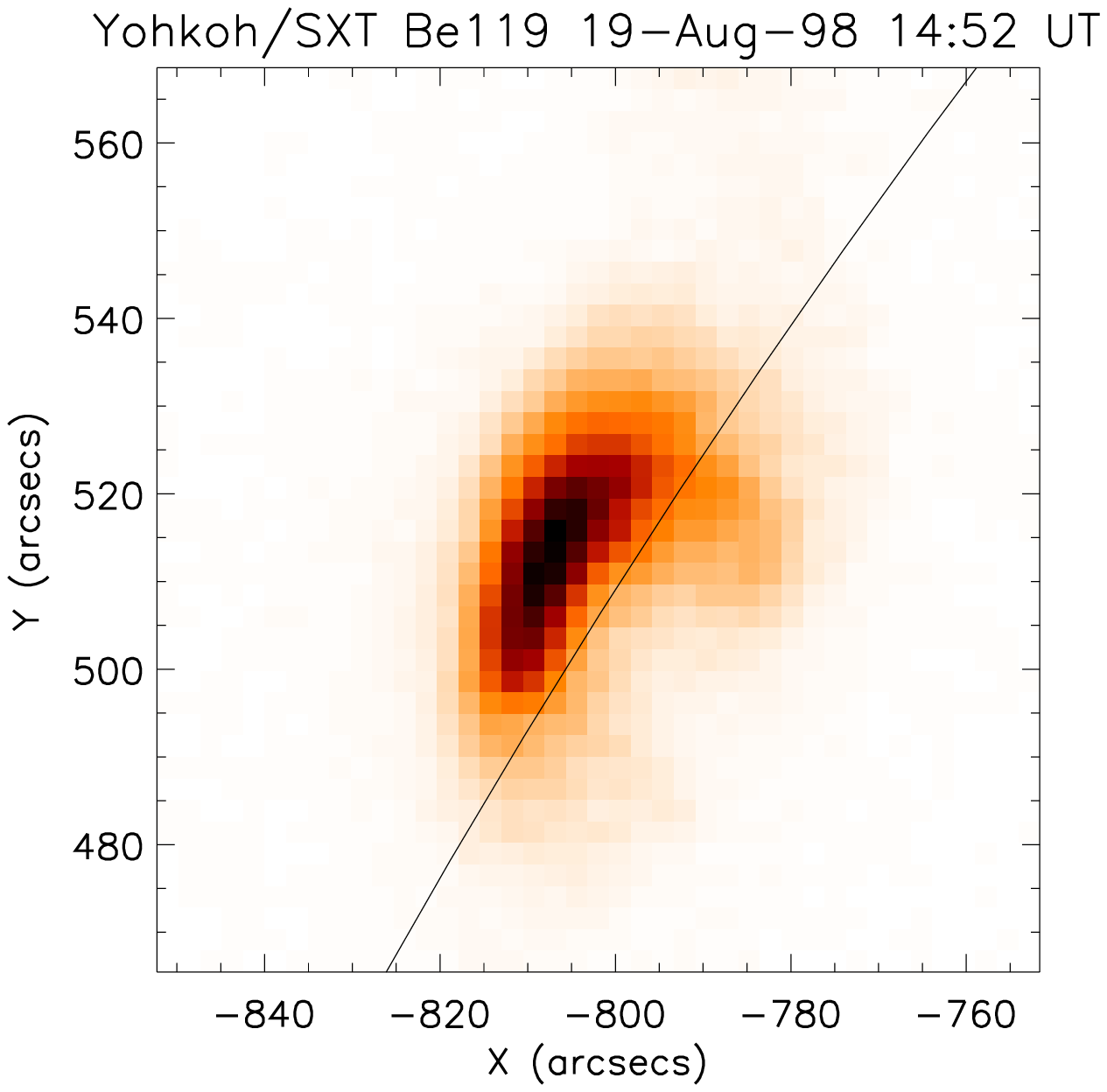}
\caption{Sample SXT(Be119) images (inverse intensity scale in colour) of August 18, 1998 ({\em left}) and August 19, 1998  ({\em right}) flares}
 \label{18_19aug98}
\end{center}
\end{figure}

Next we have calculated values of the ratio $k=N_{SXT}/N_{SS}$. The values of the ratio $k$ are given in Table \ref{table} and their distribution is shown in Figure \ref{hist}. We see that for two flares (No. $12$ and $13$) the values of $k$ are highest ($k=1.8$ and $2.2$). SXR images of these two flares are shown in Figure \ref{18_19aug98}. We see that their SXR kernels are elongated. The volumes, $V$, of such kernels were calculated under assumption that they have "sausage-like" shape {\em i.e.} that they are ellipsoids with semiaxes $a \times b \times b$. The high values of the ratio $k$ suggest that actual volume of the SXR emitting plasma is significantly larger, {\em i.~e.} the third semiaxis is larger than $b$. This indicates that estimates of $N_{SXT}$ for such elongated kernels can be loaded with highest random errors. Therefore, we have excluded these two large values of $k$ from calculation of the mean value $\bar{k}$. We have obtained:
\begin{equation}
\label{k}
\bar{k}=0.88, \hspace{0.6cm} s= \pm 0.28
\end{equation}
where $s$ is random (r. m. s.) error of single estimate of $k$.

In Figure \ref{hist} we see that $k$-values are concentrated around this mean value. The value of $\bar{k}$ which is close to one, confirms the good agreement between $N_{SXT}$ and $N_{SS}$ values. We see that random ({\em r.m.s.}) error, $s$, of $k$-values is about $30 \%$. Assuming that contribution of random errors of $N_{SXT}$ and $N_{SS}$ to this random error of $k$ are similar in size, we conclude that random error of a single determination of $N_{SXT}$ or $N_{SS}$ is about $30 \% / \sqrt{2} \approx 20 \%$. 

The values of $N_{SXT}$, obtained with the standard method, are mean electron number densities averaged over the SXR-kernel volume. But the QSS method of density determination provides actual density of the emitting plasma. Therefore, the ratio, $k=N_{SXT}/N_{SS}$, provides an estimate of the fraction of kernel's volume which is filled with the SXR emitting plasma ("filling factor"). In Figure \ref{hist} we see that $k$-values are concentrated around $k \approx 0.9$ which means that, typically, near the flare maximum the kernel's volume is nearly entirely filled with the hot SXR emitting plasma. Remaining, ($1-k$), fraction of the volume can be filled with plasma of lower temperature, $T<4$ MK, whose contribution to the investigated SXR emission is marginal. Values $k>1$ are due to random errors in both methods of density determination.

\section {Summary and conclusions}
We have found that simplified diagnostic diagrams ($T$ vs. $\sqrt{EM}$), derived from GOES observations, are adequate to identify quasi-steady-state (QSS) phase of flare evolution. Correct inclination, $\zeta\approx0.5$, of the QSS branches in the diagnostic diagrams shows that the temperature, $T_{GOES}$, derived from the GOES observations adequately represents mean temperature of the hot, SXR emitting plasma.

For the QSS branch of flare evolution we were able to determine electron number density, $N$, for SXR loop-top flare kernels, with two independent methods which are described in Sections 1 and 2. Good agreement of the $N$-values obtained with the two methods confirms reliability of the densities derived from {\em Yohkoh}/SXT images. In Section 2 we have explained why we obtain correct $N_{SXT}$ electron densities, despite of the fact that {\em Yohkoh}/SXT temperatures are systematically too low.

We have obtained that electron density near the SXR flare maximum falls within the limits of ($4-50$)$\times10^{10}$ cm$^{-3}$, but in one case of strong, compact flare it was extremely high, $N\approx8$ $\times10^{11}$ cm$^{-3}$.

Densities, $N_{SXT}$, derived from SXR images, are mean electron densities averaged over SXR flare kernels. But the densities, $N_{SS}$, obtained with the second (QSS) method, provide actual densities of emitting plasma. Therefore the ratio $N_{SXT}/N_{SS}$ provides an estimate of the filling factor of the kernel with hot, SXR emitting plasma. Our analysis of obtained $k$ values shows that, typically, flare kernels near SXR flare maximum are nearly entirely filled with the hot, SXR emitting plasma.

%
\section*{Acknowledgements} 
The {\em Yohkoh} satellite is a project of the Institute of Space and Astronautical Science of Japan. GOES is a satellite of National Oceanic and Atmospheric Administration (NOAA), USA. We thank the anonymous referee for useful comments and suggestions. This investigation has been supported by grant N203 1937 33 from the Polish Ministry of Science and High Education.

%


\end{article} 

\begin{thebibliography}{}
 
\bibitem[\protect\citeauthoryear{B\c ak-St\c e\' slicka and Jakimiec}{2005}]{bak2005}
B\c ak-St\c e\' slicka, U., Jakimiec, J., 2005,{\em Solar Phys}., \textbf{231}, 95.

\bibitem[\protect\citeauthoryear{Donnelly {\em et al.}}{1977}]{donnelly77}	                                                                             Donnelly, R. F., Grubb, R. N., Cowley, F. C., 1977, NOAA Tech. Memo. ERL SEL-48.


\bibitem[\protect\citeauthoryear{Jakimiec {\em et al.}}{1987}]{jakimiec87}
Jakimiec, J., Sylwester, B., Sylwester, J., Lemen, J. R., Mewe, R. et al.: 1987, {in: \em Solar Maximum Analysis}, Stepanov, V. E., Obridko,V. N. (Eds.), VNU Science Press, Utrecht, p. 91

\bibitem[\protect\citeauthoryear{Jakimiec {\em et al.}}{1992}]{jakimiec92}
Jakimiec, J., Sylwester, B., Sylwester, J., Serio, S., Peres, G. et al.: 1992, {\em A$\& $A} \textbf {253}, 269.

\bibitem[\protect\citeauthoryear{Ko\l oma\'nski {\em et al.}}{2002}]{kolomanski02}
Ko\l oma\'nski, S., Jakimiec, J., Tomczak, M., Falewicz, R.: 2002, {\em Adv. Space Res.} \textbf{30}, No. 3, 665.

\bibitem [\protect\citeauthoryear{Lin {\em et al.}}{2002}]{lin2002} 
Lin, R.P., Dennis, B.R., Hurford, G.J., Smith, D.M., Zehnder, A., et al.: 2002, {\em Solar Phys}. \textbf{210}, 3

\bibitem[\protect\citeauthoryear{Rosner, Tucker and Vaiana}{1978}]{rosner78}
Rosner, R., Tucker, W., Vaiana, G.: 1978, {\em Astrophys. J.} {\bf 220}, 643.

\bibitem[\protect\citeauthoryear{Tsuneta {\em et al.}}{1991}]{tsuneta91}
Tsuneta, S., Acton, L., Bruner, M., Lemen, J., Brown, W. et al.: 1991, {\em Solar Phys.} \textbf{136}, 37.
\end{thebibliography}
\end{document}